# Stretchable and self-adhesive triboelectric sensor for real-time musculoskeletal monitoring and personalized recovery

Cai Lin[1#], Yunyi Ding[2#], Kai Lin[1], Ru Wang[1], Yichen Luo[3*], Xiaofen Wu[1*]

[1]Department of Burn, The First Affiliated Hospital of Wenzhou Medical University, Nan Bai Xiang, Wenzhou 325000, China

[2]Department of Emergency Medicine, Second Affiliated Hospital, Zhejiang University School of Medicine, Hangzhou 310009, China.

[3]College of Electrical Engineering, Zhejiang University, Hangzhou 310058, China

[#]Contributed Equally to this Work.

*Corresponding author: lwuxiaofen8@wmu.edu.cn (Xiaofeng Wu), luoyichen@zju.edu.cn (Yichen Luo)

***Abstract:*** Recent advances in medical diagnostics have highlighted the importance of wearable technologies for continuous and real-time physiological monitoring. In this study, we introduce a flexible, self-powered triboelectric nanogenerator (MB-TENG) engineered from commercially available medical elastic bandages for biomechanical sensing during rehabilitation and gait analysis. Leveraging the porous and skin-friendly properties of the bandage combined with a PTFE film, the MB-TENG delivers robust electrical performance—achieving a peak open-circuit voltage ($V_{OC}$) of 122 V, a short-circuit current ($I_{SC}$) of 25 μA, and a transferred charge ($Q_{SC}$) of 110 nC—while maintaining long-term stability across 40,000 mechanical cycles. Its inherent self-adhesive property allows for multi-layer assembly without extra bonding agents, and mechanical stretching enhances output, enabling dual configurability. A stacked design further improves the power capacity, supporting applications in wearable medical electronics. The MB-TENG device seamlessly conforms to joint surfaces and foot regions, providing accurate detection of motion states and abnormal gait patterns. These features underscore the MB-TENG's potential as a low-cost, scalable platform for personalized rehabilitation, injury monitoring, and early musculoskeletal diagnosis.

## 1. Introduction

With the advancement of intelligent sensing technologies, wearable devices have become a cornerstone of modern medical monitoring systems [1–3]. Their inherent advantages—flexibility, compactness, and skin compatibility—enable applications across chronic disease management, rehabilitation, elderly care, and post-surgical monitoring [4,5]. Current platforms allow real-time tracking of vital physiological signals, such as heart rate, respiration, temperature, and blood pressure, thus supporting continuous and personalized healthcare [6,7]. Among various physiological indicators, gait patterns have drawn particular attention due to their close correlation with neuromuscular coordination, lower-limb function, and overall physical status [8]. Detailed gait analysis—encompassing force distribution, temporal sequencing, and symmetry—has been used to detect early signs of stroke-related hemiplegia, Parkinson's disease, diabetic foot complications, and post-orthopedic recovery [9,10]. Accordingly, gait monitoring has emerged as a critical frontier in wearable medical sensing. To meet clinical demands, gait sensors must exhibit high flexibility and mechanical conformity to accommodate dynamic foot deformation [11–13], while offering high sensitivity, fast response, and multi-region pressure mapping to capture fine biomechanical variations. Furthermore, real-world deployment requires low power consumption, long-term stability, and wearing comfort [14]. However, conventional pressure sensors often fail to resolve subtle gait compensations—such as asymmetrical loading due to localized pain—limiting their diagnostic utility [15,16]. These challenges underscore the need for next-generation wearable systems that can conform to diverse foot anatomies and deliver real-time, high-resolution, and non-invasive gait assessment. As the field progresses, wearable gait sensors are evolving from conceptual prototypes into clinically viable tools with the potential to enhance diagnostic precision, accelerate recovery, and enable personalized musculoskeletal care.

Since their emergence, triboelectric nanogenerators (TENGs) have been widely recognized as a promising self-powered technology in flexible electronics, owing to their ability to convert low-frequency mechanical motion into electrical signals through contact electrification and electrostatic induction [17–39]. Beyond energy harvesting, TENGs can operate as passive sensing units, offering advantages such as low power consumption, rapid response, and versatile structural adaptability [40–43], which make them particularly suitable for wearable biosignal detection systems [44–47]. In medical contexts, TENGs have demonstrated strong capabilities in capturing biomechanical and physiological signals—including gait patterns, respiration, joint motion, and cardiovascular pulses [48,49]. Their inherent mechanical compliance and efficient signal conversion enable continuous, unobtrusive monitoring—an essential feature for chronic disease management, rehabilitation, and eldercare. Compared to conventional battery-powered wearables, TENG-based devices are lightweight, autonomous, and conform well to complex body geometries, supporting long-term usage with minimal user intervention [50,51]. Despite these advantages, several barriers hinder clinical translation, particularly concerning the mechanical robustness and durability of conventional materials. For instance, commonly used metallic thin-film electrodes, while highly conductive, lack sufficient elasticity under repeated deformation and are prone to fatigue failure [52]. Alternatives such as conductive elastomers, hydrogels, and liquid metals have been explored, but they often involve complex fabrication, limited environmental stability, or high cost, which constrain scalability. To address these challenges, incorporating clinically approved soft materials into TENG designs has gained attention as a practical and translational approach [53–55]. Materials such as medical-grade elastic bandages, stretchable silicone, and hydrogel dressings offer excellent biocompatibility, breathability, and mechanical resilience. These properties not only enhance wearer comfort and long-term stability but also facilitate regulatory acceptance and integration into real-world healthcare settings. Thus, the development of wearable TENGs using flexible, skin-friendly, and clinically accepted materials represents a promising strategy toward next-generation, non-invasive, and self-powered medical sensing technologies.

In this study, we introduce a medical bandage-based triboelectric nanogenerator (MB-TENG) for continuous monitoring of gait and posture in healthcare settings. Compared to existing TENG-based wearable sensors, our approach integrates clinically approved elastic bandages to ensure biocompatibility and practical usability. The MB-TENG features dual tunability via stretching and multilayer structuring, enabling enhanced adaptability. Moreover, we demonstrate its effectiveness in real-time gait analysis and joint rehabilitation monitoring—capabilities not simultaneously realized in prior studies. By coupling a commercially sourced biocompatible bandage with a PTFE layer as the triboelectric pair, the device delivers both strong electrical performance and superior mechanical compliance. The bandage's porous and textured structure promotes effective charge separation, while offering excellent breathability and comfort for prolonged wear. Functioning in a vertical contact–separation mode, the MB-TENG achieves a maximum open-circuit voltage ($V_{OC}$) of 122 V, a short-circuit current ($I_{SC}$) of 25 μA, and a transferred charge ($Q_{SC}$) of 110 nC. The device maintains reliable output across 40,000 mechanical cycles and under various pressure conditions and frequencies. Notably, its self-adhesive layered structure and stretchable design endow it with dual tunability: strategic stretching enhances output, whereas excessive stacking may dampen performance due to reduced compressibility. To enhance output scalability, we developed a stacked MB-TENG configuration that maintains conformal skin contact while increasing energy generation, suitable for driving displays and wearable biosensors. Furthermore, the MB-TENG adapts well to curved body surfaces and enables precise, real-time monitoring of joint activity and foot motion. Its ability to detect movements in fingers, wrists, elbows, shoulders, and plantar pressure distribution underscores its clinical value for individualized rehabilitation tracking, motor recovery evaluation, and early detection of gait disorders.

## 2. Experimental Section

*2.1. Preparation of MB-TENG Device.*

The MB-TENG device was fabricated by integrating commercially available medical elastic bandage (thickness: 0.1 mm), conductive sponge (thickness: 1 mm), and PTFE film (thickness: 30 μm) into a layered structure designed for vertical contact–separation operation. The medical bandage, selected for its flexibility, breathability, and biocompatibility, served as both the substrate and one of the triboelectric layers. To construct the top electrode, a conductive sponge was first affixed onto the surface of a medical bandage strip. A second strip of bandage was then laminated over the sponge to ensure tight encapsulation and mechanical robustness. For the bottom triboelectric layer, another conductive sponge was similarly attached to a bandage strip and further covered with a PTFE film, which served as the negative triboelectric material due to its strong electron affinity. Each layer was precisely aligned and bonded to form a sandwich-like structure with the PTFE side facing the exposed bandage surface of the lower layer. Electrical wires were inserted and secured within the conductive sponge to enable signal collection, and the assembled device was sealed at the edges to prevent slippage or delamination during mechanical deformation. The incorporation of compliant and deformable materials enabled the MB-TENG to adapt to irregular body contours and withstand continuous mechanical deformation, thereby enhancing its applicability for wearable health monitoring. All constituent materials were sourced from local commercial vendors.

*2.2. Characterization and Measurements.*

The electrical and mechanical properties of the MB-TENG were characterized using a programmable vibration test system, comprising a mechanical actuator (SA-JZ050), signal controller (SA-SG030), power amplifier (SA-PA080), adjustable lift platform, and a pressure sensor (ZN5H). This setup enabled precise regulation of key actuation parameters such as motion frequency, displacement range, and contact force to mimic different mechanical stimuli. A high-sensitivity electrometer (Keithley 6517B) was employed to monitor the $V_{OC}$, $I_{SC}$, and $Q_{SC}$ in real time during operation. All tests were conducted under ambient conditions, and each dataset represents the mean value obtained from multiple measurement cycles to ensure consistency and reproducibility.

# 3. Results and discussion

*3.1. Design and fabrication strategy of the MB-TENG based on medical bandage.*

Inspired by the widespread application of elastic bandages in joint and limb support, we constructed a multifunctional triboelectric device capable of integrating seamlessly into both wearable and therapeutic settings (Fig. 1(a)). The core structural material of the device is a commercially available elastic medical bandage (Fig. 1(b)), selected for its high stretchability, skin compatibility, and breathable porous network. The device architecture follows a sandwich configuration in which the triboelectric and conductive layers are sequentially stacked and laminated (Fig. 1(c)). Specifically, conductive sponge strips are first bonded to the bandage substrate, serving as both the compliant electrode and mechanical buffer. The upper triboelectric layer incorporates a PTFE film, known for its strong electronegativity and charge retention ability, forming an effective triboelectric pair with the fibrous bandage. The assembly process follows a modular, scalable approach. The lower triboelectric unit is formed by attaching a conductive sponge onto an elastic bandage and covering it with an additional bandage layer to encapsulate the electrode (Fig. 1(d)). The top triboelectric unit is constructed similarly, with the final layer replaced by a PTFE film to enable effective contact–separation-based charge generation (Fig. 1(e)). Material photographs highlight the clinical familiarity and adaptability of the bandage components (Fig. 1(f1, f2)), while SEM images (Fig. 1(f3, f4)) reveal the dense, disordered microfiber morphology and rough surface texture of the bandage at micro- and nanoscale. These features significantly increase the effective contact area and friction during deformation, facilitating enhanced triboelectric performance. The fabricated MB-TENG exhibits excellent flexibility and a compact form factor, as shown in the physical prototypes (Fig. 1(g1, g2)). The integration of conductive sponge and PTFE components is clearly visible, with electrical wires inserted through the sponge for signal output. The overall structure is soft, conformable, and capable of withstanding repeated stretching and compression without delamination or mechanical degradation. This design enables the MB-TENG to operate efficiently under low-pressure, low-frequency biomechanical input, such as walking or joint bending, making it well-suited for self-powered health monitoring applications. Furthermore, the clinical familiarity, low cost, and body-conforming geometry of the bandage material provide a strong foundation for scalable deployment in wearable therapeutic systems and rehabilitation diagnostics.

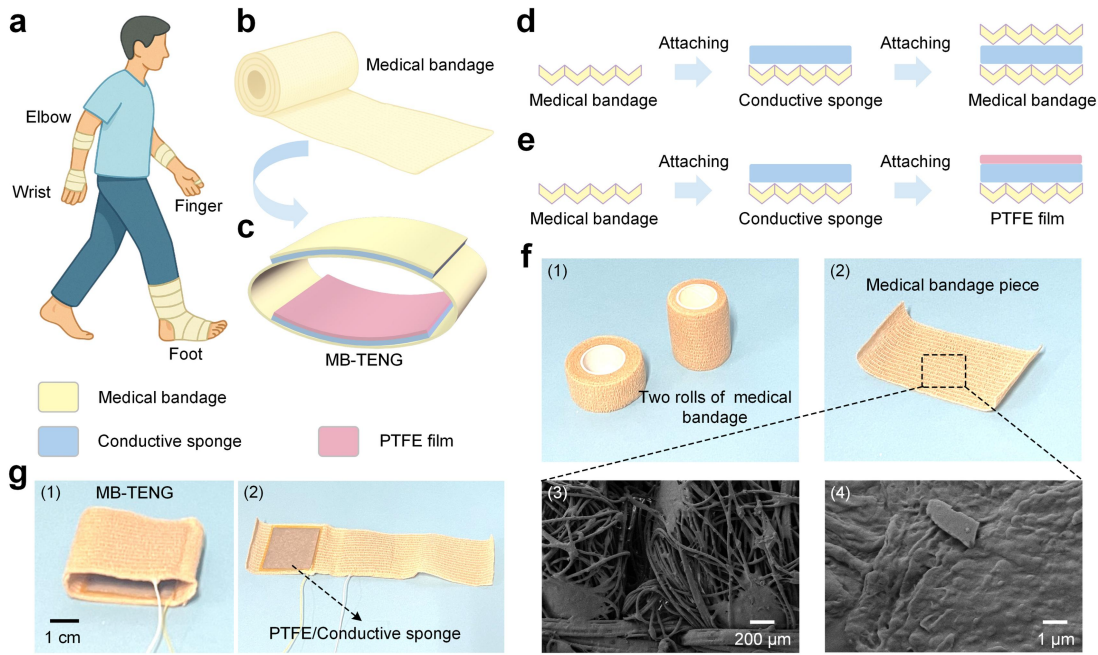

**Figure 1 | Design, structure, and fabrication of the MB-TENG based on medical elastic bandage.** (a) Schematic illustration of the MB-TENG configured for joint motion monitoring, where elastic bandages are deployed at critical articulation points (elbow, wrist, finger, foot) to detect bio-mechanical deformation and provide self-powered sensing during natural limb movements. (b) Photograph of commercially available medical elastic bandage used as the core structural and triboelectric material. (c) Exploded-view diagram of the MB-TENG structure, comprising elastic bandage, conductive sponge, and PTFE film. (d) Assembly process of the bottom layer: conductive sponge is sandwiched between two elastic bandage layers to form a flexible electrode. (e) Assembly process of the top layer: conductive sponge is attached to an elastic bandage and covered with a PTFE film to construct the triboelectric interface. (f) Physical and microscopic characterization of the medical bandage: (1) two rolls of bandage, (2) a single bandage strip, (3–4) SEM images showing the fibrous, porous microstructure and rough surface features that enhance triboelectric charge generation. (g) Photographs of the fully assembled MB-TENG: (1) folded view showing compact form factor; (2) unfolded view highlighting the inner PTFE/conductive sponge layers and wire connection.

*3.2. Working mechanism and performance tunability of the MB-TENG based on structural configuration and mechanical strain.*

The working mechanism of the MB-TENG relies on a vertical contact–separation mode that converts biomechanical motion into alternating electrical signals through the coupling of contact electrification and electrostatic induction. As shown in Fig. 2(a1), when compressive force brings the PTFE film (as the negative triboelectric material) into full contact with the medical elastic bandage (as the positive triboelectric material), electrons are transferred from the bandage to the PTFE due to their differing electron affinities. This results in the accumulation of negative charges on the PTFE surface and positive charges on the bandage, establishing an electrostatic dipole. As the force is released and the two layers begin to separate (Fig. 2(a2)), a potential difference forms between the internal electrodes, which drives electrons through the external circuit to balance the charge distribution, producing a transient current. When separation reaches its maximum (Fig. 2(a3)), the potential and current output also peak. Upon re-contact (Fig. 2(a4)), the potential difference decreases and electrons flow in the opposite direction, generating a reversed current and completing one full alternating current (AC) cycle. This contact–separation cycle repeats under continuous mechanical motion, such as joint bending or foot contact. The signal polarity and amplitude are determined by the surface charge density and separation distance, both of which can be tuned by adjusting mechanical strain or layer configuration. This enables the MB-TENG to serve as a robust, self-powered biomechanical sensor with customizable sensitivity. To investigate how device structure affects electrical output, a self-adhesive lamination approach was used to construct multilayer bandage configurations, as shown in Fig. 2(b). The medical bandage's inherent adhesion allows for layer-by-layer stacking without additional adhesives, forming tightly bonded multi-ply configurations (Fig. 2(c)). Output measurements reveal a strong dependence of device performance on the number of bandage layers. As shown in Fig. 2(d–f), the single-layer MB-TENG generates the highest output, with a peak $V_{OC}$ of 116 V, $I_{SC}$ of 31 μA, and $Q_{SC}$ of 107 nC. With increasing layers (2–4), the output exhibits a progressive decline across all metrics. This attenuation is attributed to reduced compressibility and weakened effective contact between the triboelectric surfaces. Multilayer structures introduce mechanical buffering effects that dampen the deformation amplitude, leading to suboptimal charge generation during contact cycles. In addition to layer thickness, the effect of stretching strain on triboelectric performance was systematically investigated. As illustrated in Fig. 2(g), the elastic bandage undergoes uniaxial elongation under tensile force, with minimal structural degradation. Practical stretching behavior is shown in Fig. 2(h), where the bandage is stretched from its initial relaxed state to a strain of 90%, demonstrating substantial extensibility and mechanical robustness. Fig. 2(i–k) present the output characteristics under different tensile strains (0%, 30%, 60%, and 90%). A clear enhancement trend is observed with increasing strain. The $V_{OC}$ rises from ~61 V to ~95 V, the $I_{SC}$ increases from ~3 μA to ~8 μA, and the $Q_{SC}$ grows from ~25 nC to ~55 nC. This performance enhancement arises from improved interfacial contact and frictional interaction during stretching, which facilitates more effective triboelectric charge separation and transfer. Together, these results confirm the highly tunable nature of the MB-TENG, where both layer configuration and mechanical deformation can be leveraged to optimize energy harvesting or sensing performance. While excessive stacking leads to performance degradation, controlled stretching offers a powerful route to enhance triboelectric output without compromising structural integrity. This dual-tunability strategy enables the MB-TENG to adapt to diverse wearable conditions, such as variable skin curvature or joint motion, and offers practical design flexibility for optimizing device performance in dynamic environments. The findings provide key insights into the structure–function relationship of soft triboelectric systems and lay the groundwork for tailoring TENG performance in user-specific biomedical applications.

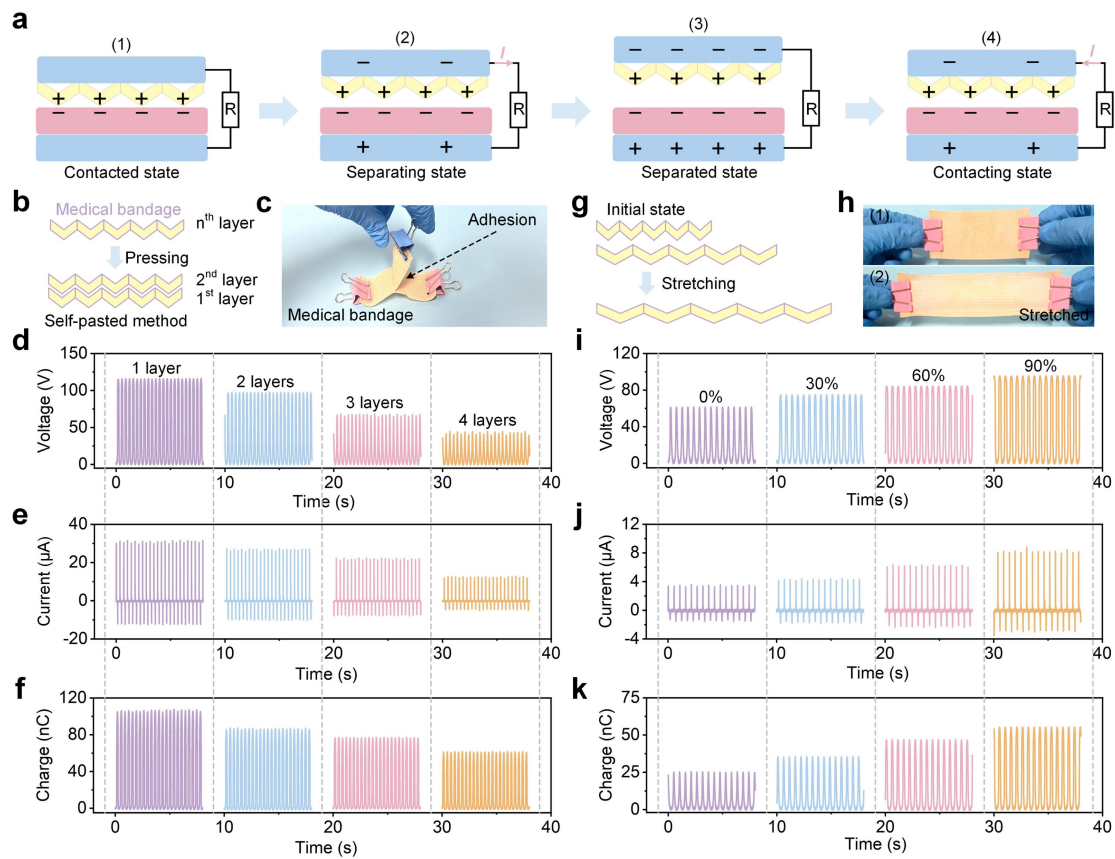

**Figure 2 | Working principle and performance tunability of the MB-TENG under structural and mechanical variations.** (a) Schematic illustration of the working mechanism of the MB-TENG in a contact–separation mode: (1) contacted state, (2) initial separation, (3) fully separated state with maximum potential difference, and (4) re-contacting state. (b) Layer-by-layer self-pasting process of medical bandage to construct multilayer triboelectric structures. (c) Photograph showing the strong adhesion between stacked bandage layers without external glue. Electrical output performance of MB-TENGs with 1 to 4 layers: (d) $V_{OC}$, (e) $I_{SC}$, and (f) $Q_{SC}$. (g)

Schematic diagram of the bandage deformation process under stretching. (h) Photographs of the bandage in the unstretched (1) and stretched (2) states, demonstrating excellent extensibility. Output signals of the MB-TENG under different stretching strains (0%, 30%, 60%, and 90%): (i) $V_{OC}$, (j) $I_{SC}$, and (k) $Q_{SC}$ output.

*3.3. Electrical performance of the MB-TENG under different mechanical conditions and its energy storage and durability characteristics.*

To assess the responsiveness and mechanical adaptability of the MB-TENG in biomechanical energy harvesting and sensing, its electrical performance was characterized under a range of dynamic conditions. The influence of excitation frequency (2–6 Hz) was first evaluated. As shown in Fig. 3(a), the $V_{OC}$ remained largely constant across the frequency range, suggesting that voltage output is primarily determined by the triboelectric interaction rather than actuation rate. However, the $I_{SC}$ shown in Fig. 3(b) increased notably with frequency—from ~5 μA at 2 Hz to over 15 μA at 6 Hz—due to shortened charge transfer intervals that accelerate electron flow. Meanwhile, the $Q_{SC}$ in Fig. 3(c) showed stable performance, indicating the reproducibility of charge output under cyclic excitation. To investigate the role of displacement, the separation distance between triboelectric layers was varied from 1 mm to 4 mm. As presented in Fig. 3(d–f), the electrical output showed a consistent rise with increasing displacement. The $V_{OC}$ increased from ~99 V at 1 mm to 122 V at 4 mm (Fig. 3(d)), while the $I_{SC}$ (Fig. 3(e)) and $Q_{SC}$ (Fig. 3(f)) peaked at 25 μA and 110 nC, respectively. This enhancement is attributed to increased contact separation, which improves charge accumulation and transfer. These observations confirm that tuning the displacement amplitude effectively boosts output performance, offering adaptive sensitivity to different biomechanical inputs such as gait intensity or joint bending. To validate energy storage functionality, the MB-TENG was coupled with a full-wave bridge rectifier and connected to capacitors of varying capacitance (Fig. 3(g)). As illustrated in Fig. 3(h), smaller capacitors (e.g., 2.2 μF) charged rapidly due to lower energy requirements, while larger ones stored more energy over extended periods. Additionally, Fig. 3(i) reveals that higher excitation frequencies (3–7 Hz) under fixed capacitance (3.3 μF) markedly enhance charging speed, indicating frequency-dependent storage efficiency. Durability was assessed by subjecting the device to 40,000 continuous operation cycles. The $I_{SC}$ output remained stable throughout (Fig. 3(j)), confirming mechanical reliability and the structural integrity of the bandage-based MB-TENG. To evaluate the structural durability of the triboelectric material, SEM characterization was conducted before and after 40,000 contact-separation cycles. As shown in **Supplementary Note 1**, the fibrous morphology of the surface remains largely unchanged, with no observable fractures or delamination. This confirms the material's mechanical robustness and long-term reliability under repetitive mechanical loading conditions. The system also responded effectively to compressive strain variations from 20% to 80% (Fig. 3(k)), with current output increasing proportionally—demonstrating the MB-TENG's strain sensitivity and applicability in pressure mapping or intensity monitoring.

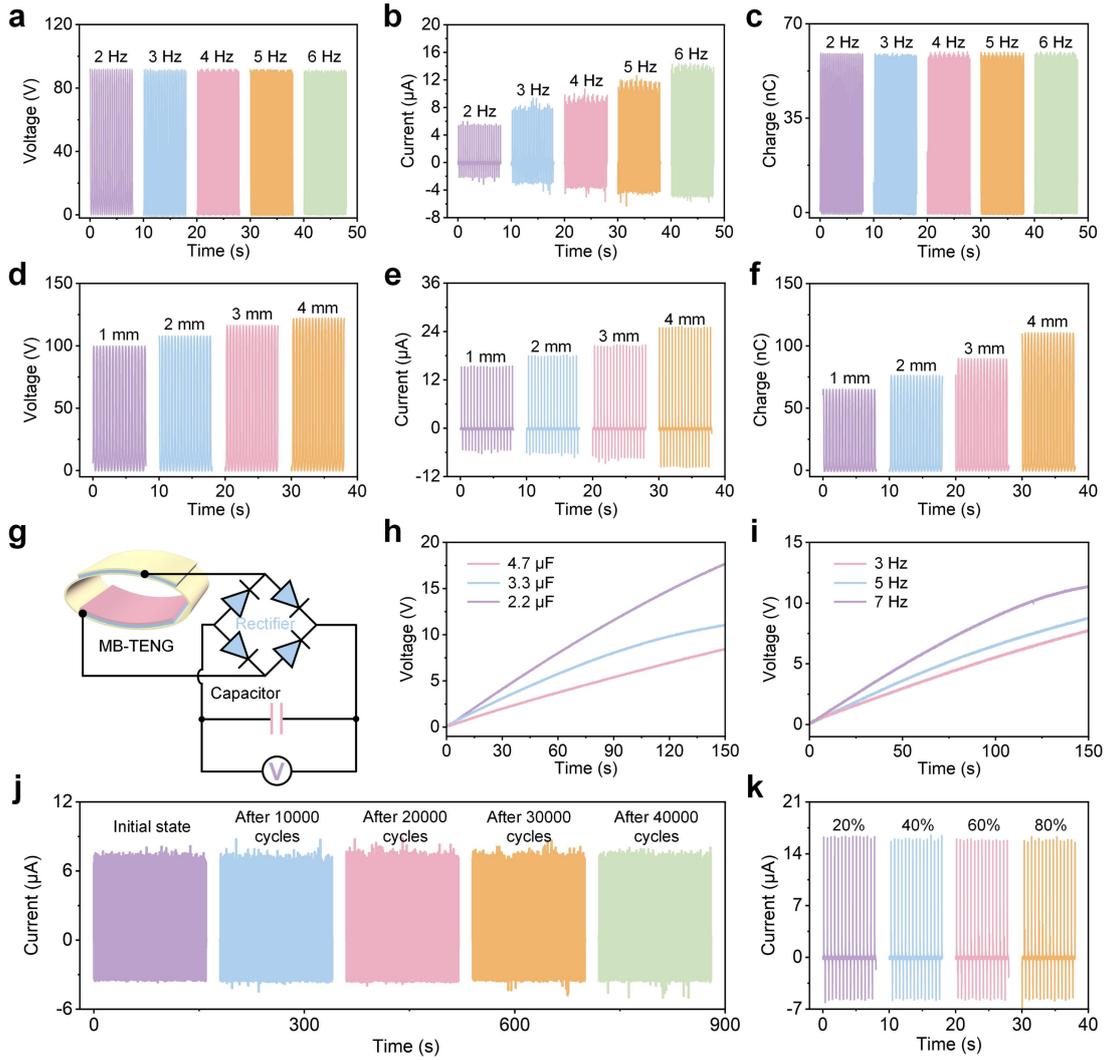

**Figure 3 | Electrical behavior of the MB-TENG under diverse mechanical stimuli, along with its energy storage capability and operational stability.** (a–c) Output signals ($V_{OC}$, $I_{SC}$, $Q_{SC}$) at excitation frequencies ranging from 2 to 6 Hz, indicating the effect of actuation rate. (d–f) Output performance under increasing vertical separation gaps (1–4 mm), including $V_{OC}$, $I_{SC}$, and $Q_{SC}$. (g) Diagram of the energy harvesting and storage circuit using a bridge rectifier connected to capacitors. (h) Charging profiles of capacitors (2.2, 3.3, and 4.7 μF) under fixed frequency, illustrating quicker voltage accumulation for smaller capacitance. (i) Charging efficiency at 3, 5, and 7 Hz using a constant 3.3 μF capacitor. (j) Long-term durability assessment showing stable $I_{SC}$ output over 40,000 continuous operation cycles. (k) Output current response to varying levels of compressive strain (20%–80%), revealing strain-dependent sensing behavior.

*3.4. Stacked MB-TENG for enhanced output performance and practical energy harvesting applications.*

To boost the electrical output of MB-TENG while maintaining its softness and suitability for wearables, a vertical stacking strategy was employed by assembling multiple MB-TENG layers. This approach enables modular power scaling and supports real-world applications such as powering sensors and small electronics. As shown in Fig. 4(a), a stacked MB-TENG is compressed by hand to simulate mechanical actuation. The layered configuration comprises two or three units connected in parallel, which work collectively to generate increased output. Electrical outputs shown in Fig. 4(b–d) demonstrate that increasing the number of MB-TENG working units significantly enhances performance: the $V_{OC}$ rises from approximately 60 V for a single unit to 100 V for a three-unit configuration; the $I_{SC}$ increases from 24 μA to 48 μA; and the $Q_{SC}$ exhibits a marked increase, though not strictly proportional, likely due to changes in internal resistance and charge redistribution during stacking. It is noted that the increase in electrical output with additional stacked units is not strictly linear. This behavior can be attributed to the reduction in overall internal impedance when multiple MB-TENGs are connected in parallel. The lower impedance leads to increased current output, but the associated redistribution of voltage and charge transfer dynamics results in a nonlinear gain in electrical output. Furthermore, mechanical constraints from stacking may slightly dampen the deformation of individual units, also affecting the scaling behavior. A corresponding circuit schematic integrating the stacked MB-TENG with a full-wave rectifier is shown in Fig. 4(e), and the capacitor charging profiles under different stacking levels in Fig. 4(f) reveal accelerated energy storage rates with more units involved. Although stacking multiple MB-TENG units increases the total transferred charge and overall energy generation, the capacitor charging curves do not exhibit strictly linear voltage growth with additional units. This behavior arises from complex interactions between internal impedance, charge redistribution, and load-capacitor matching. While the three-unit configuration delivers higher cumulative energy, the charging rate is influenced by the reduced impedance and voltage saturation effects, which can limit the final voltage attained across the capacitor. Therefore, stacking improves the charge capacity and power delivery potential, but does not guarantee proportionally faster charging or linear voltage enhancement. These results highlight the importance of optimizing load conditions and electrical architecture in multi-unit TENG systems. While vertical stacking increases the number of triboelectric interfaces and enhances total energy output, it also introduces inevitable mechanical and electrical losses. Mechanically, as more layers are added, the overall compressibility of the device decreases due to increased stiffness and reduced deformation depth per unit. This limits effective contact between the triboelectric layers and can reduce charge generation efficiency. Electrically, stacking may result in asynchronous contact–separation among units, leading to phase mismatch and partial cancellation of output signals. Additionally, variations in interfacial resistance and parasitic losses can impair charge transfer and power delivery. These factors contribute to the observed deviation from ideal output scaling and highlight the need for structural optimization when implementing multilayer configurations. In practical demonstrations, the stacked system powered a temperature-humidity sensor through hand-tapping, with voltage rising to 2.4 V (Fig. 4(g)) and sustaining repeated operation of the display unit. This showcases the MB-TENG's capacity to energize medical or environmental electronics. In this demonstration, the MB-TENG charges a 100 μF capacitor to a voltage of approximately 1.6 V, storing ~0.13 mJ of energy. This energy is sufficient to intermittently power a commercial HTC-1 temperature–humidity display module, which operates at ~1.5 V with an average current of ~20 μA (equivalent to ~30 μW power consumption). Finally, Fig. 4(h) compares the native AC output and rectified DC signal, with the latter exhibiting a smooth profile suitable for conventional devices. These findings highlight the MB-TENG's ability to harvest mechanical energy and deliver stable output through a compact, scalable, and wearable platform. Finally, the system was extended to power 60 LEDs through manual hand tapping, as illustrated in the schematic diagram (Fig. 4(i)) and the photographic demonstration (Fig. 4(j) and **Video. S1**).

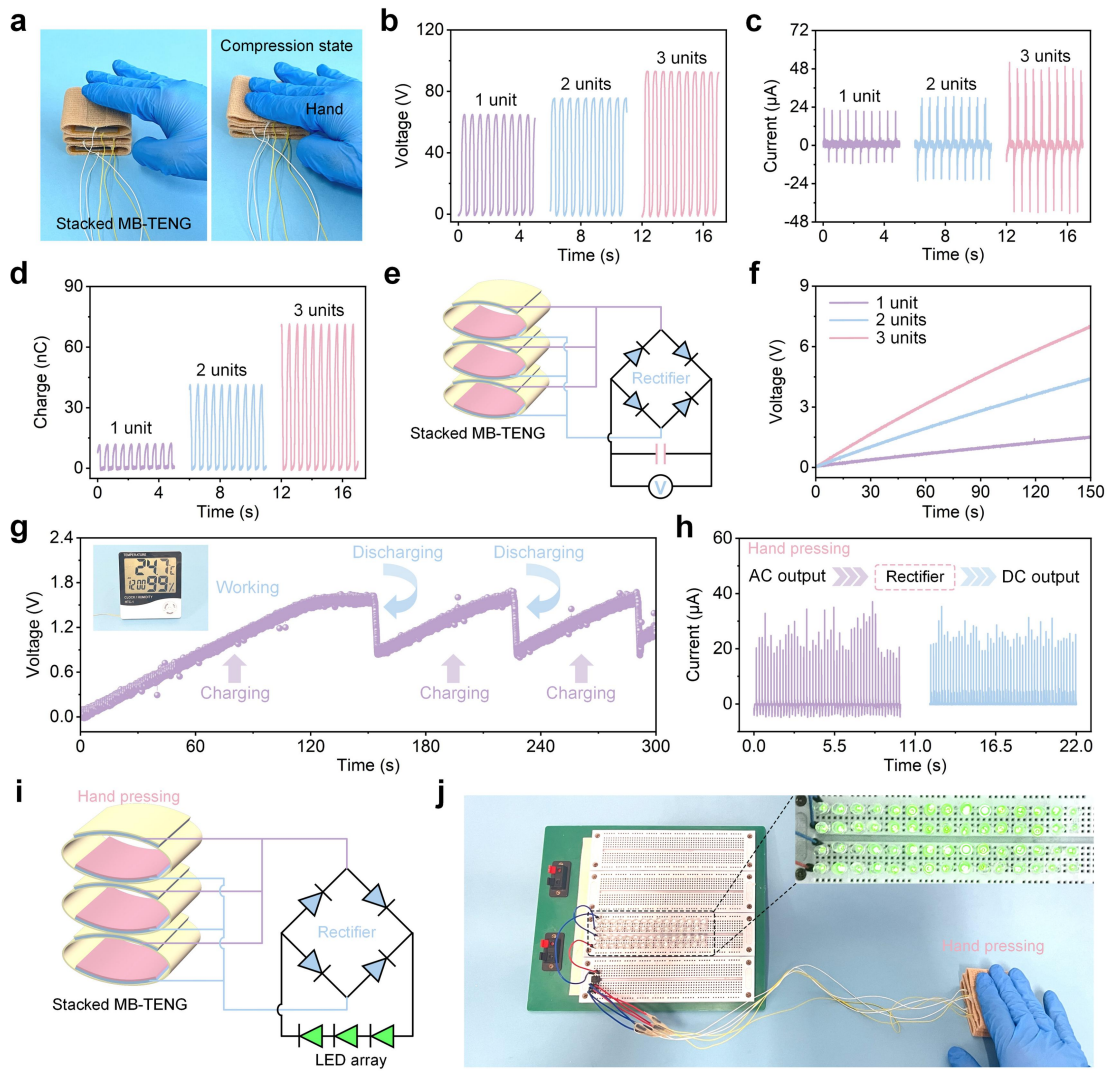

**Figure 4 | Stacked MB-TENG for enhanced electrical output and practical energy harvesting demonstrations.** (a) Photographs of the stacked MB-TENG under hand compression. (b–d) Electrical output performance of the stacked MB-TENG with 1, 2, and 3 units: (b) $V_{OC}$, (c) $I_{SC}$, and (d) $Q_{SC}$. (e) Schematic diagram of the stacked MB-TENG integrated with a full-wave rectifier for energy storage. (f) Capacitor charging curves under manual pressing for 1, 2, and 3 stacked units, indicating accelerated charging with increased output. (g) Periodic charging–discharging process powering a digital temperature–humidity sensor, with inset showing real-time operation. (h) Comparison of alternating current (AC) and rectified direct current (DC) output under hand pressing, demonstrating effective signal conversion for powering electronics. (i) Schematic of the stacked MB-TENG driving an LED array through a rectifier circuit. (j) Demonstration of real-time LED illumination powered by hand-tapping the stacked MB-TENG.

*3.5. Wearable MB-TENG for joint motion sensing and foot rehabilitation monitoring.*

    The MB-TENG was further configured into a compact, wearable drum-like structure to enable conformal integration with various human joints for motion tracking and injury assessment. This application-focused design highlights the device's potential in intelligent rehabilitation, especially for patients recovering from limb injuries or undergoing physiotherapy. Fig. 5(a) illustrates the structural design of the wearable MB-TENG, in which PTFE film, conductive sponge, and medical bandage layers are rolled into a spiral drum to form multiple triboelectric contact interfaces. This configuration ensures enhanced deformation adaptability and allows the device to wrap seamlessly around curved body regions. Fig. 5(b) introduces the envisioned clinical use case: real-time sensing of joint motion in patients with hand or foot injuries, enabling quantitative evaluation of movement range, symmetry, and recovery progress. To demonstrate localized joint monitoring, the MB-TENG was attached to finger joints, forming a soft sleeve-like sensor (Fig. 5(c)). When the finger bent

at different angles (30°, 45°, 90°), distinct voltage outputs were recorded, as shown in Fig. 5(d). The signal amplitude increased proportionally with flexion angle, validating the MB-TENG's sensitivity to joint deformation and its applicability for fine-scale range-of-motion assessment during rehabilitation exercises. Beyond individual joints, a five-channel MB-TENG array was fabricated for hand motion tracking, as shown in **Fig. S2 of Supporting Information**. Each sensor (F1–F5) was individually mounted along the back of a single finger, and its output was recorded during repeated flexion and extension of that specific finger. The resulting voltage profiles in Fig. 5(e) demonstrate that the MB-TENG can reliably detect localized joint motion with consistent waveform characteristics. To improve clarity, we have added inset schematics indicating the finger positions during bending (1) and straightening (2), allowing readers to visually correlate mechanical movement phases with corresponding signal fluctuations. Although the signals were not recorded simultaneously, the individual tests confirm the device's ability to capture fine motor activity and distinguish joint-specific movements—supporting its potential for monitoring functional limitations in neuromuscular or orthopedic conditions. To further validate the MB-TENG's performance under sustained deformation, we conducted a static flexion experiment where the finger remained in a bent position without movement for several seconds. As shown in **Fig. S3 of Supporting Information**, the voltage output maintained a stable plateau during the bending state, confirming that the device can provide consistent signals even under static conditions. This characteristic enhances its utility in rehabilitation scenarios where patients may hold positions for functional assessment or muscle endurance training. In Fig. 5(f–k), we present the voltage responses of the MB-TENG during wrist, elbow, and shoulder movements, respectively. For each joint, the flexion–extension or contraction–relaxation phases are marked along the time axis to clarify the correspondence between biomechanical motion and signal generation. The $V_{OC}$ achieved during joint flexion were approximately 5.6 V for the wrist (Fig. 5(f)), 6.8 V for the elbow (Fig. 5(g)), and 4.5 V for the shoulder (Fig. 5(h)). These values demonstrate that the MB-TENG can sensitively capture muscle-driven joint activity across multiple anatomical locations, with measurable and repeatable signal outputs. These results establish the MB-TENG as a body-conformal, self-powered sensor capable of tracking multi-joint upper-limb motion—essential for assessing post-injury performance and rehabilitation progress. Its applicability was further demonstrated in lower-limb scenarios. As shown in Fig. 5(i), the device was attached to the plantar surface to capture gait dynamics. Voltage and current signals were recorded at distinct plantar regions (P1–P3), corresponding to heel strike, midfoot stance, and forefoot lift-off (Fig. 5(j)), enabling spatial resolution of pressure distribution. Fig. 5(k) highlights its diagnostic potential: under healthy conditions, regular and strong signals were observed, while simulated foot injury produced weaker, irregular outputs, indicating asymmetrical gait. These findings demonstrate the MB-TENG's suitability for early detection of gait abnormalities, injury compensation, and post-operative monitoring. Its flexibility, biocompatibility, and continuous sensing capability offer valuable support for personalized rehabilitation and clinical musculoskeletal assessment.

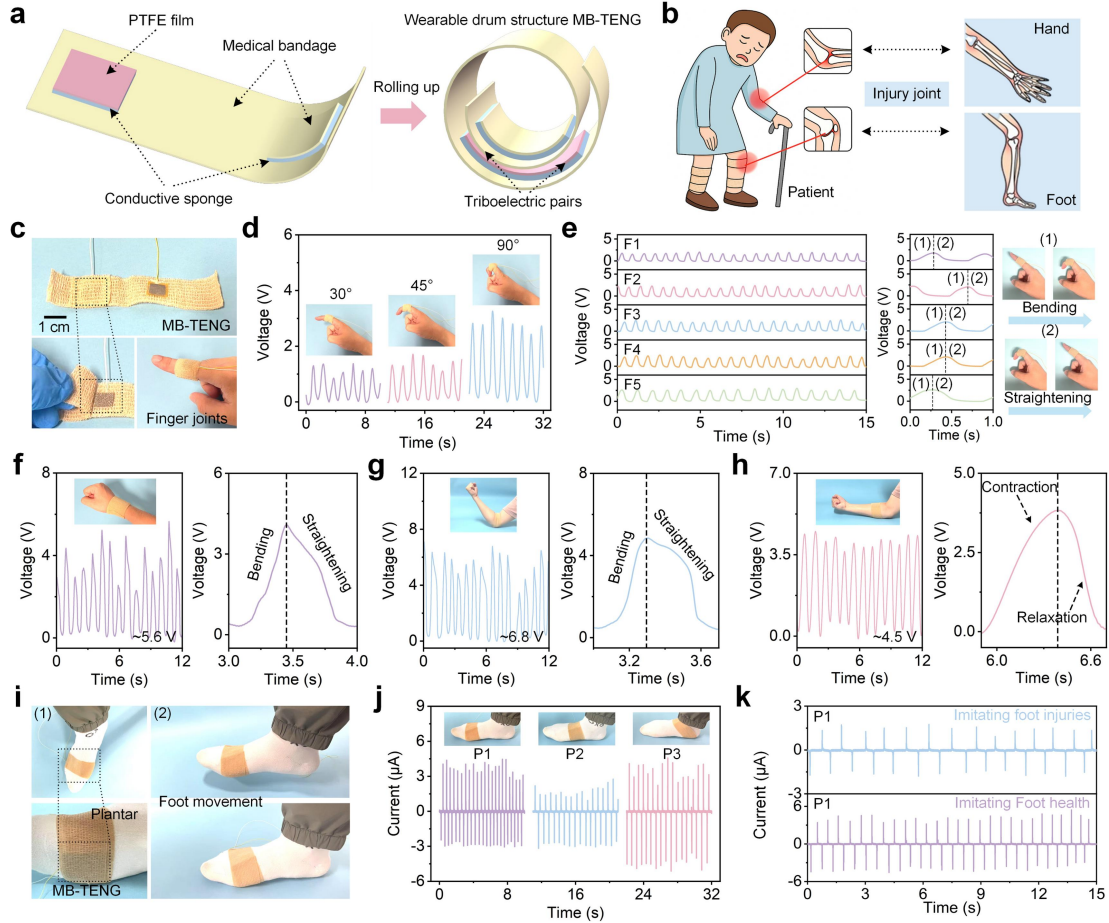

**Figure 5 | Wearable MB-TENG for joint motion sensing and foot rehabilitation monitoring.** (a) Structural design of the wearable drum-like MB-TENG composed of PTFE film, conductive sponge, and medical bandage, configured for conformal wrapping and multi-layer triboelectric contact. (b) Conceptual diagram showing application of the MB-TENG for monitoring hand and foot joints in patients with limb injuries. (c) Photograph of MB-TENG attached to finger joints using bandage-based wrapping. (d) Output voltage signals corresponding to finger bending at different angles (30°, 45°, 90°), indicating angle-dependent sensing capability. (e) Photograph of a five-channel MB-TENG array (F1–F5) applied to the back of the hand for multi-finger motion tracking. (f–k) Voltage signals from MB-TENGs placed on different upper-limb joints: (f) wrist, (g) elbow, and (h) shoulder, showing periodic signals corresponding to repeated motion. (i) Photographs of MB-TENG positioned under the foot (plantar region) to capture dynamic plantar pressure. (j) Output current responses from different foot regions (P1–P3) during walking, corresponding to heel, midfoot, and forefoot motions. (k) Comparison of current outputs between simulated healthy gait and foot injury conditions at the same plantar region (P1), demonstrating diagnostic capability for gait asymmetry and functional compensation.

## 4. Conclusions

In summary, we have developed a flexible, self-powered triboelectric nanogenerator (MB-TENG) utilizing medical elastic bandages for wearable bio-mechanical sensing in healthcare applications. The integration of a bio-compatible, porous bandage with PTFE film enables the device to achieve high electrical output while maintaining excellent stretchability, skin conformity, and long-term durability. The MB-TENG demonstrates dual tunability through its self-adhesive multilayer structure and elastic deformation, allowing output performance to be optimized for specific motion states. Moreover, a stacked MB-TENG architecture was introduced to enhance output for powering low-power electronics, confirming its scalability and versatility. The device exhibits precise and stable responses to joint bending and plantar pressure, making it suitable for real-time monitoring of gait, posture, and rehabilitation progress. This work provides a low-cost, clinically friendly strategy for constructing wearable triboelectric sensors and offers a promising platform for personalized rehabilitation, motion analysis, and early musculoskeletal diagnosis in future intelligent medical systems.

## 5. Supplementary materials

**Video S1:** Real-Time Lighting of an LED Array via Manual Tapping on the MB-TENG.

**Supplementary Note 1:** Qualitative discussion on SEM image comparability.

**Fig. S1:** SEM images of the triboelectric surface before (a) and after (b) 40,000 contact-separation cycles.

**Fig. S2:** Voltage outputs from each sensor (F1–F5) during finger flexion, demonstrating synchronized, independent joint detection.

**Fig. S3:** Voltage output of the MB-TENG during sustained finger bending.

## 6. Data Availability Statements

Data available on request from the authors.

## 7. Funding

This work was supported by the Key Science and Technology Program of Zhejiang Province (Grant No. 2023C03170).